\documentclass[]{aastex631}

\usepackage{hyperref}
\usepackage{multirow}
\newcommand{\degree}{{}$^\circ$}
\newcommand{\kms}{km s$^{-1}$}
\newcommand{\Lsun}{\rm L$_{\odot}$}

\newcommand{\water}{H$_2$O}

\shorttitle{A Black Hole Outflow-Driven Water Maser in He 2-10}
\shortauthors{Gim et al.}

\graphicspath{{./}{figures/}}

\begin{document}

\title{An Outflow-Driven Water Maser Associated with Positive Black Hole Feedback in the Dwarf Galaxy Henize 2-10}

\correspondingauthor{Hansung B. Gim}
\email{hansung.gim@montana.edu}

\author[0000-0003-1436-7658]{Hansung B. Gim}
\affiliation{Department of Physics, Montana State University, Bozeman, MT 59717, USA}

\author[0000-0001-7158-614X]{Amy E. Reines}
\affiliation{Department of Physics, Montana State University, Bozeman, MT 59717, USA}
\affiliation{eXtreme Gravity Institute, Department of Physics, Montana State University, Bozeman, MT 59717, USA}

\author[0000-0003-3168-5922]{Emmanuel Momjian}
\affiliation{National Radio Astronomy Observatory, P.O. Box O, Socorro, NM 87801, USA}

\author[0000-0003-2511-2060]{Jeremy Darling}
\affiliation{Center for Astrophysics and Space Astronomy, Department of Astrophysical and Planetary Sciences, University of Colorado, 389 UCB, Boulder, CO 80309, USA}

\begin{abstract}

Henize 2-10 is a dwarf galaxy experiencing positive black hole (BH) feedback from a radio-detected low-luminosity active galactic nucleus. Previous Green Bank Telescope (GBT) observations detected a \water\ ``kilomaser'' in Henize 2-10, but the low angular resolution ($33\arcsec$) left the location and origin of the maser ambiguous. We present new Karl G.\ Jansky Very Large Array observations of the \water\ maser line at 22.23508 GHz in Henize 2-10 with $\sim$2\arcsec\ resolution. These observations reveal two maser sources distinct in position and velocity. The first maser source is spatially coincident with the known BH outflow and the region of triggered star formation $\sim 70$ pc to the east. Combined with the broad width of the maser ($W_{\rm 50} \sim 66$ \kms), this confirms our hypothesis that part of the maser detected with the GBT is produced by the impact of the BH outflow shocking the dense molecular gas along the flow and at the interface of the eastern star-forming region. The second maser source lies to the south-east far from the central BH and has a narrow width ($W_{\rm 50} \sim 8$ \kms), suggesting a star-formation-related origin. This work has revealed the nature of the \water\ kilomaser in Henize 2-10 and illustrates the first known connection between outflow-driven \water\ masers and positive BH feedback.

\end{abstract}

\keywords{galaxies: dwarf -- galaxies: individual (He 2-10) -- galaxies: active galactic nuclei -- astrophysical masers: water masers -- ISM: molecular gas}

\section{Introduction} \label{sec:intro}

Powerful extragalactic \water\ masers with isotropic luminosities in the range $L_{\rm H_2O} \sim 10^2–10^4~L_\odot$ (the so-called ``megamasers") are found in high-density molecular gas associated with accretion disks or radio jets in active galactic nuclei \citep[AGNs; ][]{Claussen1984, Haschick1985, Braatz1994}. Megamasers with maser spots lying in a circumnuclear disk within parsecs of the central engine have yielded exceptionally accurate black hole (BH) masses as well as direct extragalactic geometric distances \citep[e.g., NGC~4258; ][]{Herrnstein1999, Kuo2011}. \water\ megamasers can also trace nuclear outflows and shocks produced by the interaction of a jet with obstructing molecular clouds \citep{Gallimore1996, Claussen1998, Wang2014}.

Extragalactic \water\ masers also manifest in a low-luminosity variety, referred to as ``kilomasers''. These moderately powerful masers ($L_{\rm H_2O} \sim 1-10~L_\odot$) are comparable to, or slightly more luminous than, the most powerful masers associated with star formation in the Milky Way \citep[e.g.\ W49N, $L_{\rm H_2O} \approx 0.15~L_\odot$; ][]{Liljestrom1989}. Consequently, extragalactic kilomasers exhibit luminosities expected for either weak AGN or strong star formation, and have indeed been associated with both using high-spatial resolution observations \citep{Henkel1986, Hagiwara2001, Tarchi2002, Brunthaler2006,Castangia2008, Darling2017, Gorski2019}. 

Henize 2-10 (He 2-10) is a blue compact dwarf galaxy that is notable for hosting a massive BH \citep{Reines2011,Reines2012,Reines2016, Riffel2020, Schutte2022} and a surrounding burst of star formation \citep{Johnson2000,Chandar2003}. 
The evidence for a massive BH in He 2-10 includes a central non-thermal radio continuum source detected with the NSF's Karl G. Jansky Very Large Array (VLA)\footnote{The National Radio Astronomy Observatory is a facility of the National Science Foundation operated under cooperative agreement by Associated Universities, Inc.} and Atacama Large Millimeter/submillimeter Array (ALMA) \citep{Reines2011,Gim2024}, a high-brightness-temperature radio core detected with the Long Baseline Array \citep[LBA; ][]{Reines2012}, and an X-ray point source detected with the {\it Chandra X-ray Observatory} \citep{Reines2016}. Moreover, \citet{Riffel2020} find an enhanced stellar velocity dispersion around this source consistent with a $1.5 \times 10^{6}$~M$_{\odot}$ BH. This is consistent with the BH mass - total stellar mass relation for AGNs \citep{Reines2015} given the stellar mass estimates in the literature for He 2-10 ($\sim 10^{9.5-10}$~M$_{\odot}$; \citealt{Reines2011,Nguyen2014}).
While the mere existence of a massive BH in such a galaxy is remarkable, the BH is also driving a low-velocity bipolar outflow that is triggering the formation of star clusters located $\sim$70 pc ($\sim$1\farcs5) away \citep{Schutte2022}. Moreover, \citet{Gim2024} provide evidence that the cold molecular gas is being shocked by the bipolar outflow from the BH, supporting the case for positive BH feedback in He 2-10. 

\begin{figure}
    \centering
    \includegraphics[angle=0,scale=1.0]{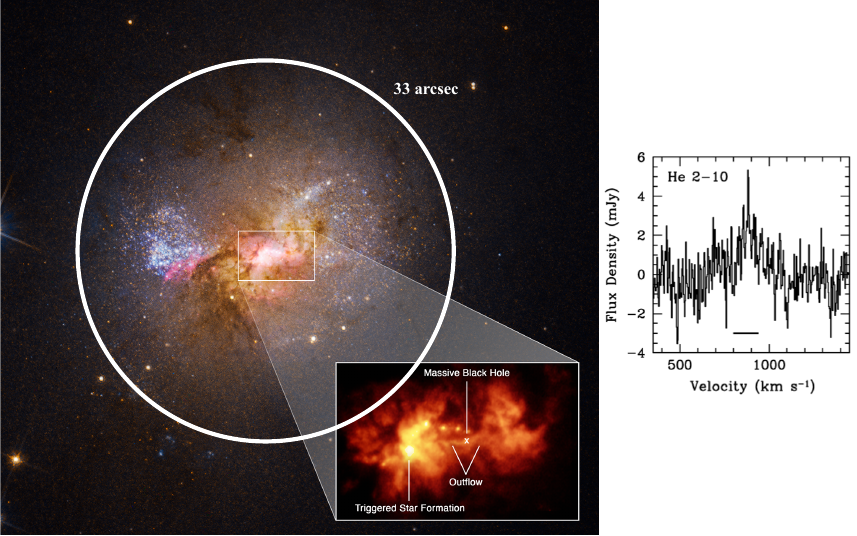}
    \caption{{\it HST} image of He 2-10 with a white circle indicating the beam size (33\arcsec) of previous GBT \water\ maser observations \citep{Darling2008}. The inset shows a narrowband H$\alpha$ image of the central 6\arcsec$\times$4\arcsec\ region, providing a detailed view of the BH outflow and region of triggered star formation \citep{Schutte2022}. The right panel displays the spectrum of the \water\ maser observed at GBT \citep{Darling2008}. Image credit: NASA, ESA, Zachary Schutte, \& Amy Reines.}
    \label{fig:he210_gbt}
\end{figure}

\water\ kilomaser has been detected toward He 2-10 in a Green Bank Telescope (GBT) survey of nearby starburst galaxies \citep{Darling2008}. 
However, the large beam of the GBT observations (33\arcsec) completely encompassed the main body of He 2-10 and left the location and origin of the maser ambiguous as shown in Figure~\ref{fig:he210_gbt}. 
The \water\ maser line detected in He 2-10 with the GBT displayed a peak flux density of 2.4$\pm$0.3~mJy and a luminosity of $L_{\rm H_{2}O} = 0.68~ L_{\odot}$. The GBT spectrum revealed a symmetric and unusually broad \water\ line with $\Delta V_{\rm FWHM} \sim$115 km s$^{-1}$ (compared to $\Delta V_{\rm FWHM} \sim$1-24 km s$^{-1}$ for the masers detected in other galaxies in the \citealt{Darling2008} sample), as shown in Figure~\ref{fig:he210_gbt}. We hypothesized that the large velocity extent of the maser line was due to the BH outflow shock-exciting nearby molecular gas analogous to jet-driven megamasers, albeit at lower power and luminosity, and proposed for high angular resolution observations to test our hypothesis.

In this paper, we report our observations of the \water\ maser in He 2-10 with the VLA with sufficient angular resolution to localize the maser and help determine its origin. We present our VLA observations in \S~\ref{sec:observations}, the results in \S~\ref{sec:results}, and give a discussion and conclusions in \S~\ref{sec:conclusions}. The distance to He 2-10 is $\sim$9~Mpc yielding a scale of $\sim 44$~pc per 1\arcsec, and the systemic velocity is 873~\kms\ \citep{Kobulnicky1995}.

\section{VLA Observations \label{sec:observations}}

Our VLA observations were conducted on 2022 Nov 1, 5, 6, 7, 11, and 12, totaling 20 hours of integration time (3 hours 20 minutes per session) in the C-configuration (project code 22B-137). Observations were carried out in the K-band (18--26.5~GHz) with 16 spectral windows (SPW) and dual polarization. Among the 16 SPWs, four had a channel width of 111~kHz (equivalent to a rest-frame velocity width of 1.5~\kms) across the 128~MHz bandwidth for spectroscopy. The remaining 12 SPWs (21.595--23.544~GHz) were used for continuum observations with a bandwidth of 128~MHz, but four SPWs had a channel width of 500~kHz and eight SPWs had a channel width of 1~MHz. We used 3C147 as the flux density scale calibrator, J0609-1542 for bandpass calibration, and J0826-2230 as the complex gain calibrator.

The calibrations and editing were carried out with the Common Astronomy Software Application \citep[CASA version 6.5.1;][]{CASA2022} using the standard calibration scheme for VLA data. No apparent radio frequency interference (RFI) was detected in our observing band. Bandpass calibration was performed by bootstrapping flux densities to account for the spectral variations across a wide bandwidth of 2.077~GHz. However, phase and amplitude errors persisted after this initial calibration process. Subsequently, we performed self-calibration to correct these issues. The central bright region was chosen for modelling the continuum emission from the source. Self-calibration was performed first in phase and then in amplitude and phase, and imaged in an iterative cycle using the continuum data of the target source. The self-calibration was done separately for each observing session.

We concatenated the six self-calibrated data sets to create the final continuum image and the water maser spectral line image cube using the CASA task \textit{tclean}. These were generated using a cell size of 0\farcs25 and image size of 256$\times$256 pixels. For the continuum image, we used multi-scale multi-term multi-frequency synthesis imaging with scales of [0, 7, 21] pixels and nterms of 3. We used a Briggs weighting function with a robust value set to 0.5 to compromise between sensitivity and synthesized beam size. The resulting synthesized beam size is 1\farcs92$\times$0\farcs88 with a position angle of -19.8\degree\ and the RMS noise is 7.9 $\mu$Jy beam$^{-1}$ for the continuum image.

For the spectral line image cube, we selected line-free channels on both sides of the emission lines by inspecting the visibilities. Continuum subtraction was performed using the CASA task \textit{uvcontsub} with a fitorder of 1. In the CASA task \textit{tclean}, the velocity of the spectral line was calculated by setting the rest-frame frequency to 22.23508~GHz with respect to the barycentric frame. The \textit{tclean} task was performed until the residual peak reached 2.5$\sigma$. We created a spectral image cube with the original channel width of 111~kHz and then smoothed the image cube to a beam size of 2\farcs05$\times$1\farcs00 with a position angle of -20.4\degree\ using \textit{imsmooth}. This spatially smoothed image cube was also smoothed spectrally through Hanning smoothing using \textit{specsmooth} in CASA to mitigate noise. Here, we applied the method of ``mean'', which yielded an RMS noise of 332 $\mu$Jy beam$^{-1}$ over a velocity resolution of 3~\kms. 

The location of the BH in He 2--10 was originally determined in the astrometric reference frame of the Two Micron All Sky Survey (2MASS), with an absolute positional accuracy of $\lesssim$ 0\farcs1. Images from {\it Hubble Space Telescope} (\textit{HST}), the VLA, and {\it Chandra} were shifted to this common reference frame in the multi-wavelength study of \citet{Reines2011}. To facilitate the comparison with other data sets and previous works, we identified the offsets in our VLA observations with respect to the 2MASS reference frame and applied them to the VLA observations. We found the offset between the position of the BH in our 22~GHz continuum image (i.e., the local maximum between the two prominent star-forming regions) and the position of the BH in the 1.4~GHz continuum image from the LBA with a synthesized beam size of 0\farcs11$\times$0\farcs03, which is given in the 2MASS absolute astrometric frame \citep{Reines2012}. The offset amounted to 0\farcs195 west in right ascension and 0\farcs13 north in declination, corresponding to 0.09 and 0.13 times the beam size of our VLA images. These computed offsets were also applied to the image cube of the \water\ maser line.

\section{Results \label{sec:results}}

\begin{figure}
    \centering
    \includegraphics[angle=0,scale=0.27]{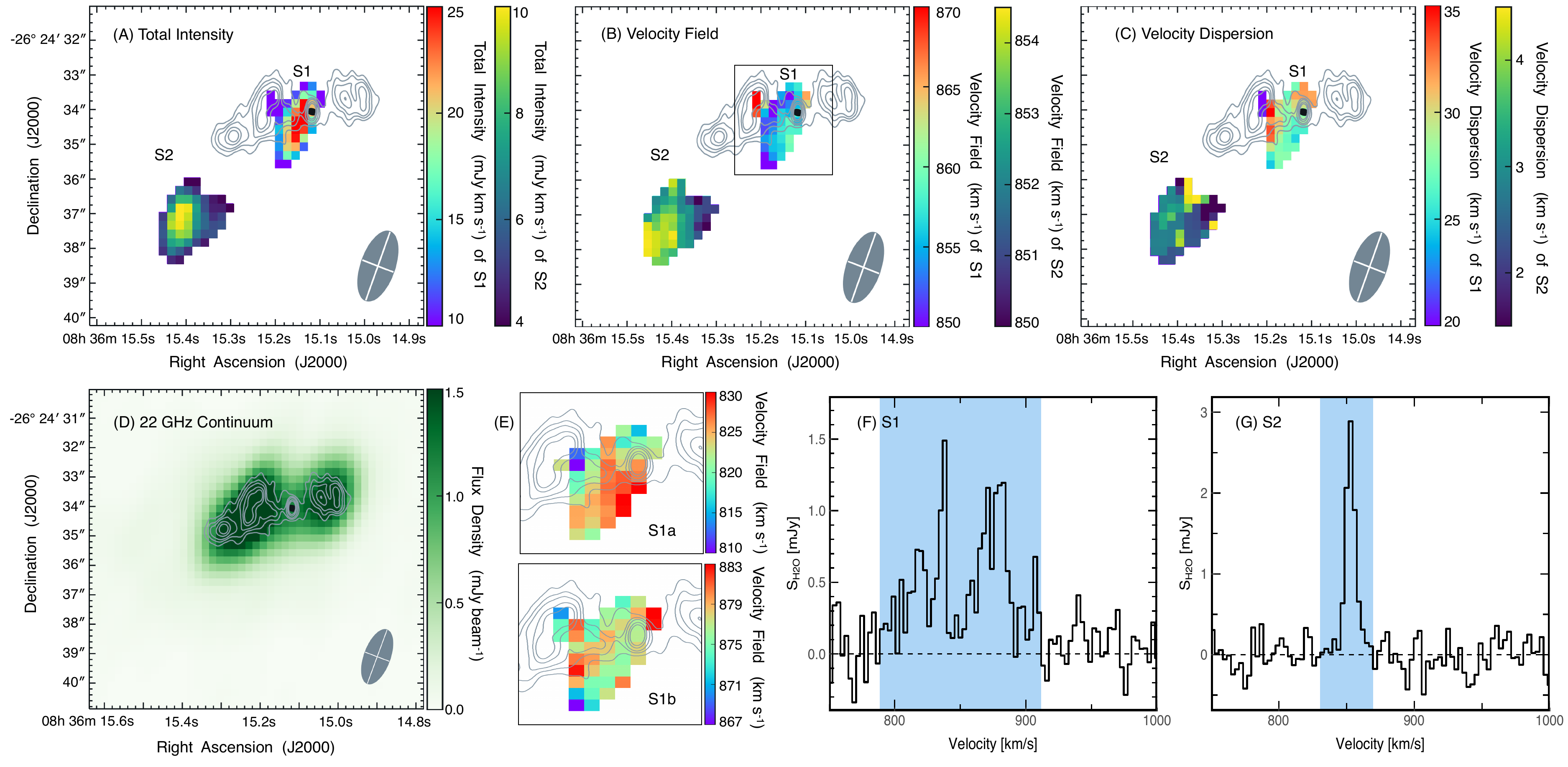}
    \caption{Panels (A-C) show moment maps of the 22.23508 GHz \water\ maser line in He 2-10. Moment 0 (total intensity), moment 1 (velocity field), and moment 2 (velocity dispersion) maps are presented in the top left, middle, and right panels, respectively. VLA X-band (8.5~GHz) contours from \citet{Reines2011} are overlaid in gray with contour levels of 9, 13, 17, 25, and 33$\sigma$, where $\sigma=12$~$\mu$Jy beam$^{-1}$. Panel (D) displays our 22~GHz continuum image of the central region of He 2-10. The massive BH is the central point source, which is also marked with a black ellipse. Panels (E) present the moment 1 maps of source S1 with two different velocity ranges, e.g., 790 -- 850 and 850 -- 913 \kms. Panels (F) and (G) show the spectra of the two maser sources, S1 and S2, detected in He 2-10. The blue shaded regions indicate the velocity width of the maser for the calculation of luminosity. The coordinates of VLA images were shifted to the 2MASS reference frame to facilitate comparison with previous multi-wavelength studies of He 2-10 \citep{Reines2011,Reines2012}.}
    \label{fig:moments}
\end{figure}

We identified two distinct regions of a \water\ maser above 3$\sigma$. The moment maps of \water\ maser lines were generated above 2$\sigma$ lines in the two sources to reflect the broad line width. Figure~\ref{fig:moments} displays these moment maps in panels (A), (B), and (C) along with contours representing the VLA X-band continuum emission, where the observations were obtained from the VLA in the A-configuration with the extension to the Pie Town antenna in the Very Long Baseline Array (VLBA), resulting the synthesized beam size of 0\farcs55$\times$0\farcs21 \citep{Reines2011}. 

The first water maser source (S1; i.e., Source 1) is located in the central region of the galaxy, between the two bright 22~GHz continuum sources (see panels A and D in Figure~\ref{fig:moments}). Specifically, S1 is positioned between the massive BH and the star-forming region to the east of the BH (see Figure~\ref{fig:moments}A). The coordinates of S1 are RA$=$08h 36m 15.148s ($\pm$0.004s) and DEC$=$-26\degree\ 24\arcmin\ 34.387\arcsec ($\pm$0.095\arcsec). The size of the source (above 3$\sigma$) has been quantified through the application of 2D Gaussian fitting utilizing the Astronomical Image Processing System (AIPS) task JMFIT \citep{Greisen2003}. This analysis, conducted on the total intensity (moment 0) map, yields 1\farcs61 ($\pm$0\farcs24)$\times$0\farcs78 ($\pm$0\farcs11), corresponding to 70 ($\pm$10)~pc$\times$34 ($\pm$5)~pc. The second maser source, S2, is outside the main central star-forming region to the south-east. Its coordinates are RA$=$08h 36m 15.402s ($\pm$0.020s) and DEC$=$-26\degree\ 24\arcmin\ 37.223\arcsec ($\pm$0.300\arcsec), and it has a size of 2\farcs08($\pm$0\farcs75)$\times$1\farcs58($\pm$0\farcs57) corresponding to 91($\pm$33)~pc$\times$69($\pm$25)~pc. The separation between S1 and S2 is 4\farcs45 (194~pc).  

Panel (B) of Figure~\ref{fig:moments} reveals that the two maser sources have distinct mean velocities: one with 847--874~\kms\ for S1 and the other with 849--854~\kms\ for S2. Spectra shown in panels (F) and (G) indicate that S1 has a wide velocity range of 790--913 \kms with $W_{\rm 50}=66.2$~\kms, while S2 has a narrow velocity range of 838 -- 865 \kms with $W_{\rm 50}=8.2$~\kms, where $W_{\rm 50}$ indicates the velocity width measured at 50\% of the highest flux density. These properties are reflected in the velocity dispersion map in panel (C), e.g., 5.6--37.2 for S1 and 1.4-5.0 \kms\ for S2. S1 has a peak flux density of 1.49$\pm$0.29~mJy (S/N$=$5.1) while S2 has a very strong line with a peak flux density of 2.89$\pm$0.29~mJy (S/N$=$9.9). We also note that the spectrum of S1 in panel (F) appears to consist of two different velocity components (S1a \& S1b). To determine if there is a spatial offset between the two components, we divided the spectrum at a velocity of 850 \kms\ and generated the moment 1 maps in panel (E). We find no positional offset through the 2D Gaussian fit. However, we note that this size of the emission is very close to the size of the beam and therefore it is difficult to separate any potential different maser regions within S1. 

The isotropic luminosities are $L_{\rm H_{2}O} = 0.11\pm 0.06 \; L_{\odot}$ for S1 and $0.05\pm 0.02~L_{\odot}$ for S2, calculated using the equation $\rm L_{\rm H_{2}O} = 0.023 D^{2} \int S dv$, where D is distance in Mpc and $\rm \int S dv$ is the total intensity in Jy \kms\ \citep{Henkel2005}. Here, the total intensities were integrated over the velocity ranges marked as blue shaded regions in panels (F) and (G) of Figure~\ref{fig:moments}. The properties of maser sources S1 and S2 are summarized in Table~\ref{tab:maser_sources}. Our estimated isotropic luminosity for the two H$_{2}$O maser sources is 0.16 $L_{\odot}$, which is a factor of 4.3 smaller than the estimate by \citet{Darling2008} using GBT observations at low angular resolution (0.68 $L_{\odot}$). This difference indicates the probable existence of multiple compact yet faint components, which may individually fall below the detection threshold of the VLA observations, yet collectively contribute to emissions detectable by the GBT. Moreover, it suggests the potential existence of an extended \water\ maser, possibly resolved out by our higher angular resolution interferometry observations, and/or variability in the maser.

\begin{table}[]
    \centering
    \begin{tabular}{cccccc}
    \hline
    \hline
     Source & Coordinate & V$_{min}$-V$_{max}$ & $W_{\rm 50}$ & S$_{peak}$ & L$_{H_{2}O}$\\
     & & (km s$^{-1}$) & (km s$^{-1}$) & (mJy beam$^{-1}$) & (L$_{\odot}$) \\
    \hline
      S1 & J083615.148-262434.387 & 790 -- 913 & 66.2 & 1.49$\pm$0.29 & 0.11$\pm$0.06 \\
    \hline
      S2 & J083615.402-262437.223 & 838 -- 856 & 8.2 & 2.89$\pm$0.29 & 0.05$\pm$0.02 \\
    \hline
    \end{tabular}
    \caption{Properties of \water\ maser sources}
    \label{tab:maser_sources}
\end{table}

\begin{figure}
    \centering
    \includegraphics[angle=0,scale=0.32]{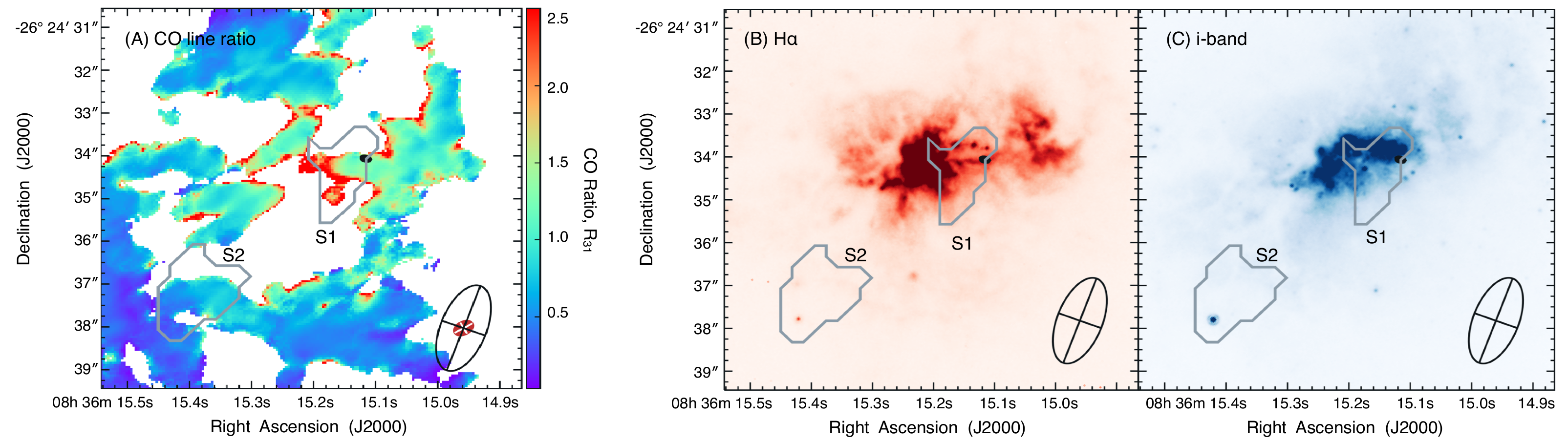}
    \caption{Panel (A) shows the CO line ratio image between CO (3--2) and CO (1--0) from \citet{Gim2024}, with the location of the \water\ maser shown as gray contours. The position of the massive BH is indicated by a black ellipse. Panels (B) and (C) show H$\alpha$ and $I$-band {\it HST} images of the same central region of the galaxy \citep{Schutte2022}. The maser source S1 extends from the BH to the region of triggered star formation to the east. It also overlaps with the interface between the BH outflow and the eastern star-forming region that exhibits a high CO line ratio due to the cold molecular gas being shocked by the BH outflow. The beam of the H$_{2}$O maser image cube is displayed in the bottom right of each panel as an open black ellipse and the beam of CO image cube is shown as filled red ellipse inside. The coordinates of VLA images were shifted to the 2MASS reference frame, facilitating direct comparison with previous multi-wavelength studies of He 2-10 \citep{Reines2011, Schutte2022, Gim2024}.}  
    \label{fig:h2o_optical}
\end{figure}

\begin{figure}
    \centering
    \includegraphics[angle=0,scale=0.7]{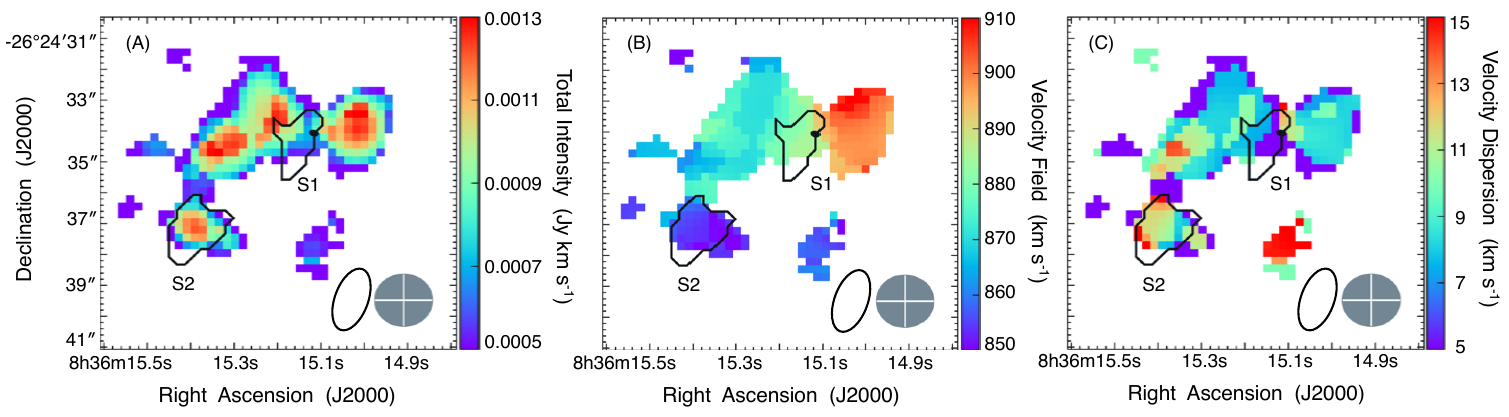}
    \caption{Moment maps of HCN (1--0) from the ALMA observations presented in \citet{Gim2024} with maser contours overlaid. Moment 0 (total intensity), moment 1 (velocity field), and moment 2 (velocity dispersion) maps are presented in the left, middle, and right panels, respectively. Black contours indicate the H$_{2}$O maser sources, and the central black filled circle marks the position of the massive BH. The beam of the HCN (1--0) image cube is displayed as a gray filled circle in the lower right corner of each panel, while the beam of the H$_{2}$O maser image cube is shown as a black open circle. The maser source S2 overlaps with an HCN-emitting cloud and has a distinct velocity from the central AGN region. This, along with the narrow width of the \water\ maser line, suggests a star-formation-related origin for S2. The coordinates of VLA images were shifted to align with the 2MASS reference frame, enabling direct comparison with prior multi-wavelength studies of He 2-10 \citep{Gim2024}.} 
    \label{fig:h2o_molecular}
\end{figure}

He 2-10 has a wealth of multi-wavelength data and we overlay the \water\ maser contours on {\it HST} and ALMA data in Figures~\ref{fig:h2o_optical} and \ref{fig:h2o_molecular}. Panel (B) in Figure \ref{fig:h2o_optical} shows that S1 is spatially coincident with the BH outflow identified by \citet{Schutte2022}, extending from the BH to the eastern star-forming region. The interface between the BH outflow and the star-forming region also exhibits a high CO line ratio between CO (3--2) and CO (1--0), which is apparent in panel (A) of Figure \ref{fig:h2o_optical} based on the ALMA observations of \citet{Gim2024}. The molecular gas mass within S1 is estimated to be $M_{\rm H_{2}} \sim $(1--8) $ \times 10^{5}$~$M_{\odot}$ using a conversion factor of $\alpha_{\rm CO}$ between 0.6 and 4.3 $M_{\odot}$ (K km s$^{-1}$ pc$^{2}$)$^{-1}$ \citep{Gim2024}. S2 is also associated with an abundance of molecular gas; most notably HCN (1--0). Figure~\ref{fig:h2o_molecular} shows the moment maps of HCN (1--0) from \citet{Gim2024} with the \water\ maser contours overlaid. S2 is spatially coincident with a blob of high-density molecular gas (traced by HCN; $n_{\rm crit} \sim 10^6$ cm$^{-3}$) with a distinct and coherent velocity structure. The associated molecular gas mass is $M_{\rm H_{2}} \sim $(3--26) $\times 10^{5}$~$M_{\odot}$. 

\section{Discussion \& Conclusions \label{sec:conclusions}}

We have presented high angular resolution VLA observations of the 22.23508 GHz \water\ maser line in the dwarf galaxy He 2-10. We detected two distinct maser sources, which were not localized in previous low angular resolution GBT \water\ maser observations of He 2-10 \citep{Darling2008}. 

The first maser source, S1, is associated with the active BH in He 2-10. S1 has an isotropic luminosity of 0.11$\pm$0.06 \Lsun\ and exhibits a size of 1\farcs61$\times$0\farcs78 (70~pc$\times$34~pc). With a 5$\sigma$ detection at its peak, S1 displays a significantly broad velocity width, implying the presence of multiple velocity components. It is located between the BH and the eastern star-forming region and coincident with a high CO line ratio, where the CO is excited by the BH outflow (see \citealt{Gim2024}). Numerous studies have reported the presence of \water\ megamasers originating from dense molecular gas shocked by an interacting BH outflow, often referred to as ``jet masers'' or ``outflow masers.'' For instance, observations of NGC~3079 conducted with the VLBA have revealed that the \water\ maser stems from a dense molecular disk subjected to interactions with an outflow within a few parsecs from the BH \citep{Trotter1998, Kondratko2005}. Similarly, NGC~1052 hosts a jet maser emitting from dense molecular clumps overlapping with the jet observed with the VLBA \citep{Braatz1994, Claussen1998}, which was supported by the consistency of a high temperature torus traced by sulfur-bearing molecular absorption lines with \water\ maser \citep{Kameno2023}. Similar phenomena have been observed in NGC~4258 through VLBA observations \citep{Haschick1994, Miyoshi1995}, NGC~1068 via VLA observations \citep{Gallimore1996, Gallimore2001} and ALMA observations \citep{Hagiwara2024}, Mrk~348 through VLBA observations \citep{Peck2003}, and the Compton-thick AGN in IRAS~F15480$-$0344 using the European VLBI Network \citep{Castangia2016}.

The preponderance of evidence supports a scenario where the \water\ maser S1 in He 2-10 is a result of the interaction between dense molecular gas and the BH outflow. However, given the low isotropic luminosity and the presence of multiple velocity components, we also consider a star-formation origin for S1. The properties of S1 are similar to the water maser emissions of the star-forming region W49N within the Milky Way Galaxy, with multiple velocity components in its spectrum and an isotropic luminosity of 0.15 $L_\odot$ \citep{Sullivan1973}. While outflow-driven \water\ masers often have a single broad line in their spectra, multiple velocity components are also seen in these objects, such as those found in NGC~1068 \citep{Hagiwara2024}, NGC~4258 \citep{Haschick1994}, and NGC~3079 \citep{Trotter1998}. Furthermore, we observe a velocity offset at S1 between HCN (1--0) and the \water\ maser. While the weighted mean velocities of HCN (1--0) range from 870 to 890 \kms\ (depicted in Figure~\ref{fig:h2o_molecular}), those of the \water\ maser range from 850 to 870 \kms\ (as shown in Figure~\ref{fig:moments}). Even considering the velocity resolution of HCN (1--0) of 10 \kms, this discrepancy in velocity field may suggest that the molecular gas is being perturbed by the BH outflow. A similar phenomenon has been identified in NGC~1068 through MERLIN and ALMA observations \citep{Hagiwara2024}. In NGC~1068, the location of the off-nuclear \water\ maser displayed velocities ranging from 1050 to 1150 \kms\ in HCN (3--2) and from 1000 to 1100 \kms\ in HCO$^{+}$ (3--2), while registering $\approx$ 980 \kms\ in the \water\ maser. Consequently, it is reasonable to infer that S1 in He 2-10 is a low power analogue of these jet/outflow \water\ megamasers.

In contrast, the second maser source S2 likely has a star-formation origin. S2 exhibits an isotropic luminosity of 0.05$\pm$0.02 \Lsun\, spanning dimensions of 2\farcs08$\times$1\farcs58 (91~pc$\times$69~pc). Situated to the southeast of S1 at a distance of 4\farcs45 ($\sim$ 194~pc) and on the periphery of the 22 GHz continuum, S2 displays a strong \water\ emission line with S/N$=$9.9 and a narrow velocity width. S2 is associated with a lower CO line ratio and dense molecular gas, indicative of its likely origin within a star-forming region. Indeed, some extragalactic kilomasers have been attributed to star formation. The investigation of three kilomasers in the Antennae galaxy using the VLA \citep{Brogan2010} demonstrated that these kilomasers were situated within the optically-obscured region and the massive molecular clouds identified via CO emission. \citet{Tarchi2011} utilized the VLA for observations of previously detected kilomasers at lower angular resolution and discovered that the kilomaser in NGC~3556 is associated with a compact radio continuum source that
is possibly a supernova remnant or radio supernova. This finding suggests that the kilomaser could potentially be generated by star formation activity. Consequently, it is highly plausible that S2 is related to star formation activity, analogous to other extragalactic \water\ kilomasers and scaled up versions of the \water\ masers in Galactic star-forming regions.

The presence of two maser sources with different origins in He 2-10 shows that \water\ maser emission can be multi-modal within a single galaxy. Moreover, this finding highlights the potential of \water\ kilomaser observations in dwarf galaxies to learn more about the influence of BHs on their host galaxies in the low-mass regime.
Our investigation also underscores the need for high spatial and spectral resolution observations of radio continuum, water masers, and molecular gas. This is necessary to trace outflows/jets from radio continuum sources, facilitate the detailed examination of molecular gas kinematics, and localize water masers. Ultimately, this will help elucidate the origins of water masers and advance our understanding of the interaction between bipolar outflows and molecular gas.

\facilities{VLA, ALMA, LBA, HST}

\software{CASA \citep{CASA2022}, CARTA \citep{CARTA}, AIPS \citep{Greisen2003}, R \citep{Rsoftware}}

\section*{Acknowledgement}
The authors express their gratitude to the anonymous reviewer for their valuable contributions, which have greatly enhanced the quality of this paper. In addition, the authors are grateful to Dr. Ilsang Yoon, Dr. Jim Braats, and Dr. Loreto Barcos Mu$\tilde{n}$oz at NRAO for helpful discussions. AER acknowledges the support provided by NASA through EPSCoR grant number 80NSSC20M0231 and the NSF through CAREER award 2235277. This research has made use of NASA's Astrophysics Data System Bibliographic Services.

\bibliographystyle{aasjournal}

\end{document}